\documentclass [12pt]{article}
\begin{document}

\title{\LARGE Monge equation of arbitrary degree in $1+1$ space}
\author{A.N.Leznov$\footnote{e-mail: andrey@buzon.uaem.mx}$\thanks{Universidad Autonoma del Estado de Morelos, CIICAp, Cuernavaca, Mexico} and R.Torres-Cordoba\thanks{Universidad Autonoma de Cd. Juarez, Chihuahua and Universidad Autonoma del Estado de Morelos, CIICAp, Cuernavaca, Mexico}}
\date{}

\maketitle

\begin{abstract}
Solution of Monge equation of arbitrary degree ($\frac{\partial^nU}{\partial x^n}=W(\frac{\partial^n U}{\partial z^n})$ is connected with solution of functional equation for 4 functions with 4 different arguments. Some number solutions of this equation is represented in explicit form.

\end{abstract}


\section{Introduction}

So called Monge-Amper equation degree $n+1$ in $1+1$ dimensional $x,z$
space looks as \cite{1}
$$
\frac{\partial}{\partial x}(\frac{\partial^n U}{\partial
x^n})\frac{\partial}{\partial z}(\frac{\partial^n U}{\partial z^n})=
\frac{\partial}{\partial y}(\frac{\partial^n U}{\partial
x^n})\frac{\partial}{\partial x}(\frac{\partial^n U}{\partial y^n})
$$
The last equation may be considered as unity to zero Jacobian between
$(\frac{\partial^n U}{\partial z^n}),\frac{\partial^n U}{\partial
x^n}$, which means their function
dependence or
$$
\frac{\partial^n U}{\partial x^n}=W(\frac{\partial^n U}{\partial z^n})
$$
The last equation we will call as Monge equation of $n$ degree with
notation $M_n$. For initial Monge Ampher equation we use term
$(M-A)_{n+1}$. Each solution of $M_n$ is simultaneous  solution of
$(M-A)_{n+1}$. General solution of $M_n$ depend on $n$ functions of
one argument and if it will be possible to find it for arbitrary $W$
function then it will be general solution for $(M-A)_{n+1}$ depending
on $n+1$ arbitrary functions of one argument.

\section{Trivial partial solution of $(M-A)_{n+1}$}

If in $M_n$ $W(x)=x$ then solution is obvious
$$
U=\sum_{k+1}^n f_k(x+\lambda_kz),\quad \lambda_k^n=1
$$
From this example it is clear that existence of the solution of $M_n$
will determine function $W$ for which solution is possible.

\section{$M_1$}
$$
U_x=W(U_z)
$$
Solution is well known connected with classical Monge equation
$\lambda_z=\lambda \lambda_x$ which can rewrite in the form of zero
Jacobian $(x+\lambda z)_x \lambda_z
-(x+\lambda z)_z \lambda_x=0$ .The last condition means functional
dependence $ x+\lambda x+\lambda z=F(\lambda)$ This is exactly general
solution in implicit form
of classical Monge equation.

\section{$\tilde M_2$}

The results of this section will be used below in sections $M_3,M_n
                      $.
Equation $\tilde M_2$ looks as
$$
U_{z,z}=W(U_{x,x},U_{z,x})
$$
Really relation above is not the equation because 3 arbitrary
functions in $1+1$ are always functionally dependent. The talk is
about some parametrization of functions
 involved in this relation.

In notations $a=U_{z,z}=W(b,c),\quad b=U_{x,z},\quad c=U_{xx}$
equation $\tilde M)_2$ looks as system of equations $ a_x= W_b b_x+W_c
c_x=b_z,\quad b_z=c_x$.

This is typical hydrodynamic system. (\cite{2}. By transformation
$x=X(b,c),z=Z(b,c)$ we calculate derivatives
$$
b_x={Z_c\over D},\quad c_x=-{Z_b\over D},\quad b_z=-{X_c\over D},\quad
c_z={X_b\over D},\quad D=Jacob(X,Z)
$$
Linear system of equation above is resolved in terms of one function
$R,X=R_c,Z=R_b$
$$
R_{cc}=-W_b R_{cb}+W_c R_{bb}
$$
There are no known methods for solution of this equation under
arbitrary function $W$. Below we propose some trick for finding
solution of this equation together
with function $W$. Let us try represent second order differential
equation in factorize form
\begin{equation}
\frac{\partial^2}{\partial c^2}+W_b\frac{\partial^2}{\partial b
\partial c}-W_c\frac{\partial^2}{\partial
b^2}=(\frac{\partial}{\partial c}+\nu_1 \frac{\partial}
{\partial b})(\frac{\partial}{\partial c}+\nu_2 \frac{\partial}{\partial b})
\end{equation}
Comparison left and right sides lead to conclusion
\begin{equation}
\nu_1+\nu_2=W_b,\quad \nu_1\nu_2=-W_c,\quad
\nu^2_c+\nu^1\nu^2_b=0,\quad \nu^1_c+\nu^2\nu^1_b=0
\end{equation}
In what follows always $\nu_2\equiv \nu^2,\quad \nu_1\equiv \nu^1$.
There are obvious three possibility in resolving the last system of
equations, which will be considered on three subsections below.

\subsubsection{$\nu_2,\nu_1=Constants$}

In this case $R=R_1(b-\nu_1c)+R_2(b-\nu_2c),\quad
W=(\nu_1+\nu_2)b-\nu_1\nu_2 c+w(R_b)$. $x=R_b=\dot R_1+R'_2,\quad
z=R_c=-\nu_1\dot R_1-\nu_2 R'_2$.
Resolving two last equations $b-\nu_2c=L_1(x+\nu_1z),\quad
b-\nu_1c=L_2(x+\nu_2z)$ or finally $ c={L_2-L_1\over
\nu_2-\nu_1},\quad c={\nu_2L_2-\nu_1L_1\over \nu_2-\nu_1}$. In the
same terms $W={\nu_2^2L_2-\nu_1^2L_1\over \nu_2-\nu_1}+w(x)$. Let us
introduce notation $l^k_s={\nu_2^s L_2^k-\nu_1^s L_1^k\over
\nu_2-\nu_1}$. $L^k_{1[2}$ $k$-derivatives of $L$ function by their
own argument. It is obvious $(l^k_s)_x=l^{k+1}_s,\quad
(l^k_s)_z=l^{k+1}_{s+1}$.

\subsubsection{$\nu_1=Constant, \nu_2=\nu_2(b-\nu_1c)$}

$\nu_2=\nu_2((b-\nu_1c)),\quad R=R_1((b-\nu_1c))+R_2(\int^{(b-\nu_1c}
d s {1\over \nu_1-\nu_2(s)}-b))$
$$
x=R_b=R_1'+\dot R_2 {\nu_1\over -\nu_2-\nu_1},\quad
z=R_c=-\nu_1R_1'+\dot R_2 {\nu_2\nu_1\over -\nu_2-\nu_1},\quad x+\nu_1
z-=\nu_1\dot R_2,
$$
$$
x+\nu_2 z=\Delta R'\equiv \Theta(\nu_2)\quad c=C(\nu_2)-L(x+\nu_1
z),\quad b=\int d\nu_2 \nu_2C_{\nu_2}-\nu_1L(x+\nu_1 z),
$$
$$
a=\int d\nu_2 \nu_2^2C_{\nu_2}-\nu_1^2L(x+\nu_1 z)+\theta(z),
$$
\subsubsection{$\nu^2_c+\nu^1\nu^2_b=0,\quad \nu^1_c+\nu^2\nu^1_b=0$}
In the system of equations in the title of this subsection let us
perform transformation $b=B(\nu^1\nu^2),\quad c=C(\nu^1\nu^2)$.
Absolutely by the same technique as above we pass to the system of
equations with obvious solution
$$
B_{\nu^2}-\nu^2 C_{\nu^2}=0,\quad B_{\nu^1}-\nu^1 C_{\nu^1}=0
$$
$$
c=C=C^1(\nu^1)+C^2(\nu^2),\quad b=B=\int d\nu^1 \nu^1 C^1_{\nu^1}+\int
d\nu^2\nu^2 C^2_{\nu^2}
$$
$$
W=\int d b (\nu_1+\nu_2)-\int dc \nu_1\nu_2+w(x)=\int d\nu^1 (\nu^1)^2
C^1_{\nu^1}+\int d\nu^2(\nu^2)^2 C^2_{\nu^2}+w(x)
$$
From equation
$$
(\frac{\partial R}{\partial c}+\nu_1 \frac{\partial R}{\partial
b})=0,\quad (\frac{\partial R}{\partial c}+\nu_2 \frac{\partial
R}{\partial b})=0
$$
 (both differential operators are commutative) it follows
$R=R^1(\nu_1)+R^2(\nu_2)$ ($\frac{\partial R}{\partial c}+\nu_1
\frac{\partial R}{\partial b})=
 (\frac{\partial R}{\partial c}-{(\nu_2)_c\over (\nu_2)_b}
\frac{\partial R}{\partial b})=0 $ from which follow result above.
Further
$$
 z=R_b=R^1_{\nu_1}(\nu_1)_b+R^2_{\nu_2}(\nu_2)_b,\quad
x=R_c=R^1_{\nu_1}(\nu_1)_c+R^2_{\nu_2}(\nu_2)_c
 $$
 or
 $$
x+\nu_1z=R^1_{\nu_1}((\nu_1)_c+{\nu_1}(\nu_1)_b),\quad
x+\nu_2z=R^2_{\nu_2}((\nu_2)_c+{\nu_2}(\nu_2)_b)
$$
After differentiation expressions above for $b,c$ functions via $\nu$
ones we obtain system of equations for determining derivatives of
$\nu$ functions
$$
\pmatrix{C^1_{\nu^1}\nu^1_c+C^2_{\nu^2}\nu^2_c \cr
      \nu^1 C^1_{\nu^1}\nu^1_c+\nu^2 C^2_{\nu^2}\nu^2_c \cr}=\pmatrix{1 \cr
0 \cr},\quad \pmatrix{C^1_{\nu^1}\nu^1_b+C^2_{\nu^2}\nu^2_b \cr
      \nu^1 C^1_{\nu^1}\nu^1_b+\nu^2 C^2_{\nu^2}\nu^2_b \cr}=\pmatrix{0 \cr
                                                                      1 \cr}
$$
Result of solution
$$
C^1_{\nu^1}\nu^1_c={\nu^2\over \nu^2-\nu^1},\quad
C^2_{\nu^2}\nu^2_c=-{\nu^1\over \nu^2-\nu^1}\quad
C^1_{\nu^1}\nu^1_b=-{1\over \nu^2-\nu^1},\quad
C^2_{\nu^2}\nu^2_b={1\over \nu^2-\nu^1}
$$
Substituting these expressions in relations connected $x,z$ variables
with $(\nu)$ one we obtain finally
$$
 x+\nu_1z={R^1_{\nu_1}((\nu_1)_c \over C^1_{\nu^1}}\equiv
\Theta^1(\nu^1),\quad x+\nu_2z={R^2_{\nu_2}((\nu_2)_c \over
C^2_{\nu^2}\equiv \Theta^2(\nu^2)}
$$
\section{$M_2$}

This case coincide with the previous one under condition $W_b=0$. The
second order differential equation looks as
$$
R_{cc}=W_c R_{bb}
$$
Solution of this equation it is possible to find in a form
 $R=\int dk e^{kb} w(k,c)$. For $w$ we pass to ordinary differential
equation of the second order $w_{cc}=k^2 W_c w$. This is typical one
dimensional Schrodinger equation with zero energy and potential
function $k^2 W_c(c)$. All cases when it is possible to find its
solution in explicit form are described in corresponding
monographies.The linear equation of second order have two fundamental
solution $w_1,w_2$ and general solution is their linear combination.
Thus
$$
W=\int d k e^{kb}( f_1(k) w_1(k)+f_2(k)w_2(k))
$$
Both fundamental solutions are depended from $k$ because potential
energy $k^2 W_c$ depends from this parameter.
Thus $(M-A)_3$ have trivial solution depending on two one dimensional
functions and series solutions connected with the cases of
integrability of corresponding ordinary differential equation also
depending on two one dimensional functions $f_1,f_2$.

\subsection{Degenerate solution}

In the main text of this section was assumed no one 2 functions from
3 ones $a,b,c$ functional  dependent. Let us consider opposite
situation.
$c=C(a)=W^{-1}(a),\quad b=B(a)$ ($W^{-1}$ inverse with respect $W$
function). From linear system of equations $a_z=b_x,\quad b_z=c_x=C_a
a_x$ we immediately obtain $
B_a^2=C_a$ and the first equation is usual one dimensional Monge
equation $a_z=C_a^{{1\over 2}}a_x$ with solution in implicit form
$x+C_a^{{1\over 2}}z=G(a)$. Thus we have solution in degenerate case
$M_2,(M-A)_3$ equations depending on 2 arbitrary functions
$(W^{-1}(a),G(a))$.

\section{$M_3$}
$$
U_{xxx}=W(U_{zzz})
$$
This equation as in section possible system form in notations
$a=U_{z,z,z},\quad b=U_{x,z,z},\quad c=U_{x,x,z}\quad d=U_{x,x,x}$.
$d_z=c_x=W(a)_z,\quad a_z=b_x,\quad b_x=c_z$. In two dimensional all
three functions are functionally depended. Thus it is possible
represented $a=f(b,c)$
The system of equations under such assumption take form
$$
f_b b_x+f_c c_x=b_z,\quad b_x=c_z,\quad c_x=W_b b_z+W_c c_z,\quad W=W(f)
$$
After the same transformation as in section $\tilde M_2$ we pass to
two equations in partial derivatives which must  be self consistent
\begin{equation}
f_b R_{b,c}-f_c R_{b,b}= -R_{c,c},\quad -R_{b,b}=-W_b R_{c,c}+W_c
R_{c,b},\quad W=W(f),\quad x=R_c,\quad z=R_b
\end{equation}

\subsection{$\nu_2,\nu_1=Constants$}

In this case $R=R_1(b-\nu_1c)+R_2(b-\nu_2c),\quad
W=(\nu_1+\nu_2)b-\nu_1\nu_2 c+w(R_b)$. $x=R_b=\dot R_1+R'_2,\quad
z=R_c=-\nu_1\dot R_1-\nu_2 R'_2$.
Resolving two last equations $b-\nu_2c=L_1(x+\nu_1z),\quad
b-\nu_1c=L_2(x+\nu_2z)$ or finally $ c={L_2-L_1\over
\nu_2-\nu_1},\quad b={\nu_2L_2-\nu_1L_1\over \nu_2-\nu_1}$. In the
same terms $f={\nu_2^2L_2-\nu_1^2L_1\over \nu_2-\nu_1}+\sigma(x)$
For determination $W$ we have additional equations
\begin{equation}
\nu_1^{-1}+\nu_2^{-1}=W_c,\quad (\nu_1\nu_2)^{-1}=-W_b,\quad
\nu^2_c+\nu^1\nu^2_b=0,\quad \nu^1_c+\nu^2\nu^1_b=0
\end{equation}
$$
W=c(\nu_1^{-1}+\nu_2^{-1})-b
(\nu_1\nu_2)^{-1}+\theta(z)={\nu_2^{-1}L_2-\nu_1^{-1}L_1\over
\nu_2-\nu_1}+\theta(z)
$$
Thus in notation of the section $\tilde M_2$ we have
$$
c=l^0_0,\quad b=l^0_1,\quad a=f=l^0_2+\theta(z),W=l^0_{-1}+\sigma(x)
$$
But $W=W(f)$. This fact is equivalent as equality to zero Jacobian
between this functions $Jacbian (f,W)=0$ with arbitrary two arguments
of the problem. Choosing
$x,z$ as such arguments have
$$
f_x W_z=f_z W_x,\quad (l^1_3+\theta_z)(l^1_{-1}x+\sigma_x)=l^1_2 l^1_0
$$
or
\begin{equation}
\sigma_x \theta_z)+\sigma_xl^1_3+\theta_z l^1_{-1}+{[]\over
\nu_1\nu_2} \dot L_2 L'_1=0
\end{equation}
In what follows following notations are used
$$
\Delta=\nu_2-\nu_1,\quad []=(\nu_1^2+\nu_1\nu_2+\nu_2^2),\quad
\rho={\nu_1+\nu_2)\nu_1\nu_2\over []}
$$
Equation above is functional equation connected 4 functions $\dot
L_2,L_1',\sigma_x,\theta_z$ with 4 different arguments $
x+\nu_2z,x+\nu_1z, x,z$.
This functional equation may be rewritten in many different forms. We
present one of them from which we will be able to obtain some number
of partial solutions
\begin{equation}
{\theta_z+ \nu_1^2 L_1'\over \nu_1^{-1}L_1'-\nu_2\sigma_x}={\theta_z+
\nu_2^2 \dot L_2\over \nu_2^{-1}\dot L_2-\nu_1 \sigma_x}
\end{equation}
Both functional equations are invariant with respect to the following
transformation
$$
\sigma_x\to {[]\over \nu_1^4\nu_2^4 \sigma_x},\quad \theta_z\to
{[]\over \theta_z},\quad L_1'\to {1\over \nu_1^2 L_1'},\quad \dot
L_2\to {1\over \nu_1^2 \dot L_2}
$$
We would like to show that if one derivatives is constant functional
equation have some explicit partial solution.

\subsubsection{$\sigma_x=A=Constant$}

In this case let us introduce in the last equation notation
$u^{-1}(x+\nu_1z)=\nu_1^{-1}L_1'-\nu_2 A,\quad v^{-1}(x+\nu_2
z)=\nu_2^{-1}\dot L_2-nu_1 A$ and determine $\theta_z$ in this terms.
Result is the following one
$$
\theta_z={\nu_2^3-\nu_1^3\over u-v}+{\nu_2^3\nu_1 A u-\nu_2\nu_1^3 A v\over u-v}
$$
Derivative with respect to $x$ the last expression equal to zero and
$u_x=u',v_x=\dot v$
The final result
$$
{u'\over u+(\tilde A)^{-1}}={\dot v\over (v+(\tilde A)^{-1}}=k,\quad
\ln(v+(\tilde A)^{-1}) =k(x+\nu_2z)+d_2\equiv kD_2,\quad \tilde
A\equiv \rho A
$$
$$
v^{-1}= (e^{kD_2}-(\tilde A)^{-1})^{-1}=\nu_2^{-1}\dot
L_2-\nu_1A,\quad u^{-1}=(e^{k D_1}-1)^{-1}=\nu_1^{-1}L_1'-\nu_2 A,
$$
$$
\theta_z=A\nu_2\nu_1{\nu_2^2 e^{kD_1}-\nu_1^2 e^{D_2}\over e^{kD_1}-e^{kD_2}}
$$
Trivial integration of the last expressions lead to
$$
L_2={\nu_2\tilde A\over k}\ln(1-\tilde A e^{-kD_2})+\nu_2\nu_1
AD_2,\quad L_1={\nu_1\tilde A\over k}\ln(1-\tilde Ae^{D_1})+\nu_2\nu_1
AD_2
$$
$$
\theta=-\tilde A{\nu_2^3-\nu_1^3\over k \Delta}\ln
(e^{-kD_2}-e^{-kD_1})-A\nu_2\nu_1 [(\nu_2+\nu_1)x+[] z]  \quad
\sigma=Ax
$$
$$
W=l^0_{-1}+\sigma=A{\nu_1 D_2-\nu_2 D_1\over \Delta}+{\tilde A\over
k}\ln{1-(\tilde A)^{-1} e^{-kD_2}\over 1-\tilde Ae^{-kD_1}}+Ax={\tilde
A\over k}\ln{1-(\tilde A)^{-1} e^{-kD_2}\over 1-(\tilde
A)^{-1}e^{-kD_1}}
$$
$$
f=a=l^0_2+\theta=A{\nu_2 ^2D_2-\nu_1 ^2D_1)\over \Delta}+
\nu_2^3{\tilde A\over \Delta k}\ln(1-\tilde A e^{-kD_2})-
$$
$$
\nu_1^3{\tilde A\over \Delta k}\ln(1-\tilde A e^{-kD_1})-\tilde
A{\nu_2^3-\nu_1^3\over k \Delta}\ln (e^{-kD_2}-e^{-kD_1})-A\nu_2\nu_1
[(\nu_2+\nu_1)x+[] z]=
$$
$$
\nu_2^3{\tilde A\over \Delta k}\ln{(1-\tilde A
e^{-kD_2})\over(e^{-kD_2}-e^{-kD_1}} -
\nu_1^3{\tilde A\over \Delta k}\ln{(1-\tilde A e^{-kD_1}\over
(e^{-kD_2}-e^{-kD_1}})
$$
or
\begin{equation}
{\Delta k\over \tilde A} f=(\nu_2^3-\nu_1^3)\ln(e^{W {k\over \tilde
A}}-1)+\nu_1^3{W k\over \tilde A}
\end{equation}

\subsubsection{$L_1'=D=Constant$}

In this case let us rewrite the main equation in equivalent form
\begin{equation}
{\nu_2^{-1}\dot L_2-\nu_1 \sigma_x L_1'\over
\nu_1^{-1}L_1'-\nu_2\sigma_x}={\theta_z+ \nu_2^2 \dot L_2\over
\theta_z+ \nu_1^2 L_1'}
\end{equation}
and introduce notations $u^{-1}
u^{-1}(x)=\nu_1^{-1}D-\nu_2\sigma_x,\quad v^{-1}(z)=\theta_z+
\nu_1^2D$. In this notations the last equation resolved with respect
to $\dot L_2$ looks as
$$
\dot L_2={1-{\nu_1\over \nu_2}\over
\nu_2^{-1}u-\nu_2^2v}+D{\nu_2^{-1}u-\nu_1^2v\over
\nu_2^{-1}u-\nu_2^2v}
$$
After action on the this expression by operator
$H=\frac{\partial}{\partial z}-\nu_2\frac{\partial}{\partial x}$
keeping in mind that $HL_2=0,\quad Hu=-\nu_2u_x,\quad Hv=v_z$ we pass
to equation for $u,v$  functions with obvious solution ($\tilde
D=D(\nu_+\nu_2)$)
$$
{u_x\over \nu_2^2\tilde D^{-1}+u}+\nu_2^{-1}{v_z\over \nu_2^{-1}\tilde
D^{-1}+v}=0,\quad \nu_2^{-1}\tilde D^{-1}+v=e^{-k\nu_2z},\quad
\nu_2^2\tilde D^{-1}+u=e^{kx}
$$
Substituting these expressions ion equation for $\dot L_2$ and using
definitions of $u,v$ functions we obtain explicit expressions for all
derivations
$$
\dot L_2=D+D(\nu_2^2-\nu_1^2){e^{-kD_2}\over
\nu_2^{-1}-\nu_2^2e^{-kD_2}},\quad
\sigma_x=\nu_1^{-1}\nu_2^{-1}D-\nu_2^{-1}{1\over e^{kx}-\nu_2^2\tilde
D^{-1}}
$$
$$
L_1'=D,\quad \theta_z=-\nu_1^2D+{1\over e^{-\nu_2kz}-\nu_2^{-1}\tilde D^{-1}}
$$
Trivial integration leads to explicit expressions of
$L_2,L_1,\theta,\sigma$ functions
$$
L_2={\nu_1^2\over \nu_2^2}D x+\nu_2D z+D{\nu_2^2-\nu_1^2\over
k\nu_2^2}\ln (\nu_2^{-1}e^{kx}-\nu_2^2e^{-k\nu_2z}),\quad
L_1=D(x+\nu_1z)
$$
$$
\theta=-D[]z-D{\nu_2+\nu_1\over k}D\ln (e^{-k\nu_2z}-\nu_2^{-1}\tilde
D^{-1}),\quad \sigma=\nu_2^{-3}\nu_1^{-1}[]Dx-\nu_2^{-3}{\tilde D\over
k}\ln (e^{kx}-\nu_2^2\tilde D^{-1})
$$
With help of these formulae and definitions above we have
$$
W=l^0_{-1}+\sigma=\nu_2^{-3}{\tilde D\over
k}\ln(\nu_2^{-1}-\nu_2^2{e^{-k\nu_2z}-\nu_2^{-1}\tilde D^{-1}\over
e^{kx}-\nu_2^{-1}\tilde D^{-1}}),
$$
$$
 a=f=l^0_2+\theta={\tilde D\over
k}\ln(\nu_2^{-1}{e^{kx}-\nu_2^{-1}\tilde D^{-1}\over
e^{-k\nu_2z}-\nu_2^{-1}D^{-1}}-\nu_2^2)
$$
or functions $W,f$ are functionally dependent
\begin{equation}
\nu_2^{-1}e^{{-k\nu_2^3W\over \tilde D}}-\nu_2^2e^{{-kf\over \tilde D}}=1
\end{equation}
\subsubsection{$\theta_z=E=Constant$}

After introduction new notation ${1\over u}(x+\nu_1z)=E+ \nu_1^2
L_1',\quad {1\over v}(x+\nu_2z)=E+ \nu_2^2 \dot L_2$ and resolving the
main equation (\ref{10} with
respect To$\sigma_x$ we have
$$
\sigma_x(\nu_1v-\nu_2u=\nu_2^{-3}-\nu_1^{-3}+E({\nu_1^{-3}u-\nu_2^{-3}v})
$$
After calculation derivative with respect to $z$ argument and trivial
manipulations as in previous sub subsections we pass to equations foe
$u,v$ functions with obvious
solution
$$
{\nu_1 u'\over u+{\nu_1\over E \rho}}=k={\nu_2\dot v\over
v+{\nu_2\over E\rho}}\quad u=e^{{k\over \nu_1}D_1}-{\nu_1\over E
\rho},\quad
v=e^{{k\over \nu_2}D_2}-{\nu_2\over E \rho}
$$
From definitions of $u,v$ functions above $\sigma_x$ via these
functions we obtain
$$
\nu_2^2 \dot L_2=-E+{1\over e^{{k\over \nu_2}D_2}-{\nu_2\over E
\rho}},\quad \nu_1^2 L_1'=-E+{1\over e^{{k\over \nu_1}D_1}-{\nu_1\over
E \rho}},\quad
\sigma_x=E{\nu_1^{-3}e^{{k\over \nu_1}D_1}-\nu_1^{-3}e^{{k\over
\nu_2}D_2}\over \nu_1e^{{k\over \nu_2}D_2}-\nu_2e^{{k\over \nu_1}D_1}}
$$
Result of integration
$$
\nu_2^2 L_2=-E D_2 +{\rho\over k}\ln(1-{\nu_2\over E\rho} e^{-{k\over
\nu_2}D_2}),\quad \nu_1^2 L_1=-E D_1 +{\rho\over k}\ln(1-{\nu_1\over
E\rho}
e^{-{k\over \nu_1}D_1})
$$
$$
\sigma=-\nu_1^{-3}\nu_2^{-1}-{\nu_1+\nu_1\over
\nu_2^3\nu_1^3}D_2-{\nu_1+\nu_2\over
k\nu_2^2\nu_1^2}(\equiv{[]\rho\over k\nu_2^3\nu_1^3})
\ln (\nu_1e^{{-k\over \nu_1}D_1}-\nu_2e^{-{k\over \nu_2}D_2})
$$
$$
f={\rho\over k\Delta}\ln{(1-{\nu_2\over E\rho} e^{-{k\over
\nu_2}D_2})\over (1-{\nu_1\over E\rho}e^{-{k\over \nu_1}D_1})},\quad
W={\rho\over k\Delta}[\nu_2^{-3}\ln(1-{\nu_2\over E\rho} e^{-{k\over
\nu_2}D_2})-\nu_1^{-3}\ln(1-{\nu_1\over E\rho} e^{-{k\over
\nu_1}D_1})-
$$
$$
(\nu_2^{-3}-\nu_1^{-3})\ln (\nu_1e^{-{k\over
\nu_1}D_1}-\nu_2e^{-{k\over \nu_2}D_2})].
$$
From the last two relations above we obtain
$$
-{\rho\over k\Delta}W=\nu_2^{-3}\ln (1-e^{f{\rho\over
k\Delta}})-\nu_1^{-3}\ln (e^{-f{\rho\over k\Delta}}-1)
$$

\subsection{$\nu_1=Constant, \nu_2=\nu_2(b-\nu_1c)$}

$\nu_2=\nu_2((b-\nu_1c)),\quad R=R_1((b-\nu_1c))+R_2(\int^{(b-\nu_1c}
d s {1\over \nu_1-\nu_2(s)}-b))$
$$
x=R_b=R_1'+\dot R_2 {\nu_1\over -\nu_2-\nu_1},\quad
z=R_c=-\nu_1R_1'+\dot R_2 {\nu_2\nu_1\over -\nu_2-\nu_1},\quad x+\nu_1
z-=\nu_1\dot R_2,
$$
$$
x+\nu_2 z=\Delta R'\equiv \Theta(\nu_2)\quad c=C(\nu_2)-L(x+\nu_1
z),\quad b=\int d\nu_2 \nu_2C_{\nu_2}-\nu_1L(x+\nu_1 z),
$$
$$
a=\int d\nu_2 \nu_2^2C_{\nu_2}-\nu_1^2L(x+\nu_1 z)+\theta(z),\quad
W=\int d\nu_2 \nu_2^{-1}C_{\nu_2}\nu_1^{-1}L(x+\nu_1 z)+\sigma(x)
$$
and only one equation remains
$$
W=W(a),\quad \sigma_x\theta_z+\theta_z(\nu_2^{-1}C_{\nu_2}(\nu_2)_x-\nu_1^{-1}L')+\sigma_x(\nu_2^3C_{\nu_2}(\nu_2)_x-\nu_1^3L')=
{\Delta^2[]\over \nu_1\nu_2}(\nu_2)_xC_{\nu_2}L'.
$$
also functional equation of the same kind as above and below ones.

\subsubsection{$\theta_z=E=constant$}

Only with the aim to demonstrate self consistence of this equation
let us consider the simplest case $L'=0$ and simplest solution
$\nu^2=-{x\over z}, \Theta =0,\nu^2_x=-{1\over z},\nu^2_z={x\over
z^2}$, Functional equation reduced
 to
$$
{1\over C_{\nu_2}(\nu_2)}+{1\over x \sigma_x}+{1\over
z^4\theta_z}x^3=0,\quad {1\over \nu_2^2} ({1\over
C_{\nu_2}(\nu_2)})_{\nu_2}+({1\over z^4\theta_z})_zz^4=0
$$
The second equation arises after differentiation the first one with
respect to $z$. Solution of this system is the following one
$$
C_{\nu_2}=k\nu_2^3+\alpha,\quad \theta=-{1\over 3k}\ln (k
z^{-3}+\beta) ,\quad \sigma={1\over 3\alpha}\ln (\alpha x^{-3}+k)
$$
After simple calculations we obtain functional dependence $W,a=f$
functions in a form
$e^{3\alpha W}+{\alpha\over k} e^{-3k f}={\beta\over k}$.

\subsection{$\nu^2_c+\nu^1\nu^2_b=0,\quad \nu^1_c+\nu^2\nu^1_b=0$}

We present only finally formulae of the corresponding subsubsection of
section 4.
$$
c=C^1_{\nu^1}+C^2_{\nu^2},\quad b=\int d \nu^1 \nu^1 C^1_{\nu^1}+\int
d \nu^2 \nu^2 C^2_{\nu^2}
$$
$$
f=\int[(\nu_1+\nu_2)dc-\nu_1\nu_2db]+\theta(z)=\int d \nu_1 \nu_1^2
C^1_{\nu_1}+\int d \nu_2 \nu_2^2 C^2_{\nu_2}+\theta,\quad
(\nu^i=\nu_i)
$$

$$
W=\int [(\nu_1^{-1}+\nu_2^{-1})db-(\nu_1\nu_2)^{-1}dc]+\sigma(x)=\int
d \nu_1 \nu_1^{-1} C^1_{\nu_1}+\int d \nu_2 \nu_2^{-1}
C^2_{\nu_2}+\sigma
$$
Substituting these expressions in relations connected $x,z$ variables
with $(\nu)$ ones we obtain finally
$$
 z+\nu_1x \Theta^1(\nu^1),\quad z+\nu_2x=\Theta^2(\nu^2),\quad
z={\Theta^2-\Theta^1\over \Delta},\quad
={\nu_1\Theta^2-\nu_2\Theta^1\over \Delta}
$$
Thus only one problem which remains to resolve is functionally
dependence of $f,W$ functions or $f_zW_x=f_xW_z$. Taking into account
equations defined
$\nu$ functions we have
$$
\nu^1_x={1\over \Theta^1_{\nu_1}-z},\quad \nu^1_z={\nu^1\over
\Theta^1_{\nu_1}-z},\quad \nu^2_x={1\over \Theta^2_{\nu_2}-z},\quad
\nu^2_z={\nu^2\over \Theta^2_{\nu_2}-z}
$$
\begin{equation}
\Delta \theta_z\sigma_x+(\nu_1^3 C^1_{\nu_1}{\Delta\over q_1}+\nu_2^2
C^2_{\nu_2}{\Delta\over q_2})\sigma_x+(\nu_1^{-1}
C^1_{\nu_1}{\Delta\over q_1}+ \nu_2^{-1} C^2_{\nu_2}
{\Delta\over q_2})\theta_z=
$$
$$
({\nu_1^2\over \nu_1}-{\nu_2^2\over \nu_1}){\Delta\over
q_1}{\Delta\over q_2}  C^1_{\nu_1}C^2_{\nu_2},\quad
q_1=\Theta^1_{\nu_1}-z,\quad q_1=\Theta^2_{\nu_2}-z
\end{equation}
which can be rewritten in equivalent form
$$
{\nu_2^{-1} C^2_{\nu_2}{\Delta\over q_2}-\nu_1\sigma_x\over
-\nu_1^{-1} C^1_{\nu_1}{\Delta\over q_1}-\nu_2\sigma_x}=
{\theta_z+\nu_2^2 C^2_{\nu_2}{\Delta\over q_2}\over \theta_z-\nu_1^2
C^1_{\nu_1}{\Delta\over q_1}}
$$
an thus (\ref{10}) and last above is functional equation connected 4
functions $ C^1,C^2,\theta,\sigma$ with corresponding arguments
$(\nu^1,\nu^2,z,x)$.

Of cause we have no idea how to find general solution of (\ref{10})
and we present below one of its partial solution to prove its self
consistence.

\subsubsection{$\theta_z=E=Constant$}

Absolutely by the same way as in previous subsubsection we assume
$\theta_z=A=Const$ and choose $\Theta$ functions in special form
$\Theta^1=\nu_1^2,\quad \Theta^2=\nu_2^2$ for which $
z=\nu_1+\nu_2,\quad x=-\nu_1\nu_2,\quad q_2=\Delta,\quad q_1=-\Delta$.
 Under such restrictions we resolve
equation (\ref{10}) with respect to $\sigma$
function in a form
\begin{equation}
\sigma_x={{\nu_1^{-1}C^1_{\nu_1}(E+\nu_2^2C^2_{\nu_2})-\nu_2^{-1}C^2_{\nu_2}(E+\nu_1^2}C^1_{\nu_1})\over
\nu_2(E+\nu_2^2C^2_{\nu_2})-
\nu_1(E+\nu_1^2C^1_{\nu_1})}=-x^{-1}{P-\bar P\over Q-\bar Q}
\end{equation}
where $P(\nu_1)={C^1_{\nu_1}\over \nu_1(E+\nu_1^2C^1_{\nu_1}},\quad
\bar P(\nu_2)={C^2_{\nu_2}\over \nu_2(E+\nu_2^2C^2_{\nu_2}},\quad
Q={1\over \nu_1(E+\nu_1^2C^1_{\nu_1}},
\quad \bar Q={1\over \nu_2(E+\nu_2^2C^2_{\nu_2})},\quad  \nu_1^2
P+EQ={1\over \nu_1},\quad
\nu_2^2 \bar P+E\bar Q={1\over \nu_2}$
Let us choose $Q=a \nu_1+E^{-1}\nu_1^{-1},\quad P=-aD\nu_1^{-1},\quad
\bar Q=a \nu_2+E^{-1}\nu_2^{-1},\quad \bar P=-aD\nu_2^{-1}$. Under
such choice
we have
$$
x \sigma_x={E\over x+(aD)^{-1}},\quad C^1_{\nu_1}=-{E\over \nu_1^2
+(aD)^{-1}}, \quad C^1_{\nu_1}=-{E\over \nu_1^2 +(aD)^{-1}},\quad
\theta_z=E
$$
From main equation it follows that it is invariant with respect to
multiplication all unknown functions $\theta,\sigma ,C^2,C^1$ on
common factor. By this reason we will omit
common factor $E$ in solution above and denote $(aD)^{-1}=g$. Now in
connection with beginning  of section 6 we calculate
$$
W=-\int d \nu_1 {1\over (\nu_1^2 +g)\nu_1}-\int d \nu_2 {1\over
(\nu_2^2 +g)\nu_2}+\sigma={1\over 2g}\ln {(\nu_2^2 +g)(\nu_1^2
+g)\over \nu_2^2 \nu_1^2}+
$$
$$
{1\over g}\ln {x\over x+g}={1\over 2g}\ln(1+g{z^2\over (x+g)^2})
,\quad z=\nu_2+\nu_1,\quad x=-\nu_2 \nu_1
$$
$$
f=-\int d \nu_1 {\nu_1^2\over (\nu_1^2 +g)}-\int d \nu_2 {\nu_1^2\over
(\nu_1^2 +g)}+\theta={1\over 2\sqrt{-g}}\ln{1+\sqrt{-g}{z\over
(x+g)}\over 1-\sqrt{-g}{z\over (x+g)}}
$$
Thus in this case solution of $M_3$ exists if function $W$ as function
of its argument $f$ looks as
$$
e^{2gW}={1\over \cosh^2(f\sqrt{-g})}
$$
There are other solution if assume from the beginning that $E=0$. Then
equation (\ref{11}) takes the form
\begin{equation}
\sigma_x={({\nu_2^2\over \nu_1}-{\nu_1^2\over
\nu_2})C^2_{\nu_2}C^1_{\nu_1}\over
\nu_2^3C^2_{\nu_2}-\nu_1^3C^1_{\nu_1}}=-x^{-1}{\nu_1^{-3}-\nu_2^{-3}\over
Q-\bar Q}
\end{equation}
where $Q={1\over \nu_1^3C^1_{\nu_1}}\quad \bar Q={1\over
\nu_1^3C^1_{\nu_1}}$. Resolving of (\ref{12})
$$
 Q=a \nu_1^3+b+c\nu_1^{-3},\quad \bar Q=Q=a
\nu_2^3+b+c\nu_2^{-3},\quad \sigma_x=-{1\over x(ax^3+c)}
$$
$$
\sigma={1\over 3c}\ln(a+c x^{-3}),\quad C^1_{\nu_1}={1\over a}{1\over
(\nu_1^3+\alpha_1)(\nu_1^3+\alpha_2)},\quad
C^2_{\nu_2}={1\over a}{1\over (\nu_2^3+\alpha_1)(\nu_2^3+\alpha_2)}
$$
where $\alpha_1+\alpha_2={b\over a},\quad \alpha_1\alpha_2={c\over a}$.
$$
f=\int d \nu_1 {\nu_1^2\over
a(\nu_1^3+\alpha_1)(\nu_1^3+\alpha_2)}+(\nu_1\to \nu_2)={1\over
3a(\alpha_2-\alpha_1)}
\ln{(\nu_1^3+\alpha_1)(\nu_2^3+\alpha_1)\over
(\nu_1^3+\alpha_2)(\nu_2^3+\alpha_2)}\equiv
$$
$$
{1\over 3a(\alpha_2-\alpha_1)} \ln{Q_1\over Q_2},\quad \alpha_2Q_1-
\alpha_2Q_2=(\alpha_1-\alpha_2)(x^3+\alpha_1\alpha_2)
$$
$$
W=\int d \nu_1 {\nu_1^{-1}\over
a(\nu_1^3+\alpha_1)(\nu_1^3+\alpha_2)}+(\nu_1\to \nu_2)+\Delta={1\over
3a(\alpha_2\alpha_1)}(\ln \nu_1^3\nu_2^3)-
$$
$$
{\alpha_2\over \alpha_2-\alpha_1}\ln Q_1+{\alpha_1\over
\alpha_2-\alpha_1}\ln Q_2)+{1\over 3c}\ln {\alpha_2Q_1-
\alpha_1Q_2\over (\alpha_1-\alpha_2)x^3}=
$$
$$
{\alpha_2\over 3c( \alpha_2-\alpha_1)}\ln {\alpha_2- \alpha_1{Q_2\over
Q_1}\over (\alpha_1-\alpha_2)}-{\alpha_1\over 3c( \alpha_2-\alpha_1)}
\ln {\alpha_2{Q_1\over Q_2}- \alpha_1\over (\alpha_1-\alpha_2)}
$$
And finally dependence of $W$ from $f$ under which solution of $M_3$
exists in this case is the following one
\begin{equation}
3c( \alpha_2-\alpha_1)W=\alpha_2\ln {\alpha_2- \alpha_1
e^{-3a(\alpha_2-\alpha_1)f}\over (\alpha_1-\alpha_2)}-\alpha_1\ln
{\alpha_2 e^{3a(\alpha_2-\alpha_1)f}-
 \alpha_1\over (\alpha_1-\alpha_2)}
\end{equation}

\section{$M_n$}

The results of previous sections may be generalized on the case of
$M_n$ equations with arbitrary $n$.

Let us introduce notations $a^k=(\frac{\partial^n U}
{\partial z^{n-k}}{\partial z^k}\quad k=0,1,...n-1$, then $M_n$
equation $a^n=W(a^0)$   may be represented in a form
$$
a^k_x=a^{k+1}_z,\quad a^{n-1}_x=a^n_z=W(a^0)_z=W_{a^0} a^0_z
$$
Now let us  use parametrization by one of 3 possibilities of previous
sections. For instance $\nu_1,\nu_2=constants$ and functions $l_k^s$
are defined by the
same way as above. There are not difficult to check that functions
$a^k$ defined as (all notations see in section $\tilde M_2$)
$$
a^0=l^0_{n-1}+\theta (z),\quad a^1=l^0_{n-2},\quad,
a^2=l^0_{n-3}......  \quad  W=l^0_{-1}+\sigma(x)
$$
satisfy all equations above except of condition of functionally
dependence functions $W+\sigma_x$ and $a^0+\theta(z)$
\begin{equation}
(l^1_n+\theta_z)(l^1_{-1}x+\sigma_x)=l^1_{n-1} l^1_0,\quad \sigma_x
\theta_z+\sigma_xl^1_n+\theta_z l^1_{-1}+{[]_n\over \nu_1\nu_2} \dot
L_2 L'_1=0
\end{equation}
where $[]_n\equiv {\nu_2^n-\nu_1^n\over \nu_2-\nu_1}$.
We emphasize that among solution of last equation the trivial one is
exists. Really
let $\theta=\sigma=0$ and $W=a^0$ then last condition means that (we
assume also that $L_1=0$) $({1-\nu_2^n\over \nu_2}\dot L_2=0$ and this
is exactly trivial solution of
section 2.  It is understand that solution in the considerable case
$M_n$ equation may be obtained from case $M_3$ by very simple
manipulations.
The example below clarify situation.

\subsubsection{$\theta_z=E=Constant$}

Equation (\ref{14}) in form resolved with respect to $\sigma_x$ in notations
$$
 \bar R(D_2)={\dot L_2\over \nu_2(\nu_2^{n-1}\dot L_2+E)},\quad \bar
Q(D_2)={1\over \nu_2(\nu_2^{n-1}\dot L_2+E)},\quad \nu_2^{n-1}\bar
R+E\bar Q={1\over \nu_2}
$$
$$
R(D_1)={L'_1\over \nu_2(\nu_2^{n-1}L'_1+E)},\quad Q(D_1)={1\over
\nu_1(\nu_1^{n-1}L'_1+E)},\quad \nu_1^{n-1} R+E Q={1\over \nu_1}
$$
looks as
$$
\sigma_x\nu_1\nu_2=-{\bar R-R\over \bar Q-Q}
$$
After differentiation the last with respect to $z$ we pass to
$$
0=(\nu_2\dot {\bar R}- \nu_1 R')(\bar Q-Q)-(\bar R-R)(\nu_2\dot {\bar
Q}- \nu_1 Q')
$$
Substituting $Q,\bar Q$ via $R,\bar R$ we obtain equations for
determining $R,\bar R$ functions
$$
{1\over \nu_1}({1\over R'}+\nu_1\nu_2[]_{n-1}{R\over R'})={1\over
\nu_2}({1\over \dot {\bar R}}+\nu_1\nu_2[]_{n-1}{\bar R\over \dot
{\bar R}})={1\over k}=Constant
$$
with solution
$$
R=-{1\over \nu_1\nu_2[]_{n-1}}+ce^{k\nu_2[]_{n-1} D_1},\quad \bar
R=-{1\over \nu_1\nu_2[]_{n-1}}+\bar ce^{k\nu_1[]_{n-1} D_2}
$$
$$
EQ={[]_n\over \nu_1\nu_2[]_{n-1}}-\nu_1^{n-1}ce^{k\nu_2[]_{n-1}
D_1},\quad E\bar Q={[]_n\over \nu_1\nu_2[]_{n-1}}-\nu_2^{n-1}\bar
ce^{k\nu_1[]_{n-1} D_2}
$$
Substituting these results in equation for $\sigma_x\nu_1\nu_2$
$$
\sigma_x\nu_1\nu_2==E{\bar ce^{k\nu_1[]_{n-1} D_2}-ce^{k\nu_2[]_{n-1}
D_1}\over \nu_2^{n-1}\bar ce^{k\nu_1[]_{n-1}
D_2}-\nu_1^{n-1}ce^{k\nu_2[]_{n-1} D_1}}
$$
and all other calculations as in the case of $M_3$ equation. Explicit
expressions for $Q,\bar Q$ define $\dot L_2,L_1'$ . After integration
these expressions we
obtain explicit form $\sigma L_2,L_1,\theta=Ez$ functions some partial
solution of $M_n$ equation  and explicit form of $W$ function under
which this solution
exists.

\section{Conclusion remarks}

In the present paper was proposed and used the following construction.
Infinite dimensional chain of of equations $ a^k_x=a^{k+1}_z$ was
realized on the space
of 2 one   dimensional functions in 3 version $ \nu_{1,2}^k
L_{1,2}(x+\nu_{1,2}z),\nu_{1,2}=Constant,\quad \nu_1^k
L(x+\nu_1z),\int d\nu_2 \nu_2^kC^2_{\nu_2},
\nu_1=Constant,x+\nu_2 z=\Theta(\nu_2),\quad \int d\nu_1 \nu_1^k
C^1_{\nu_1},\int d\nu_2 \nu_2^kC^2_{\nu_2},x+\nu_1
z=\Theta(\nu_1),x+\nu_2 z=\Theta(\nu_2)$.
After this this chain was interrupted by additional condition $
a^{n+1}+\sigma(x)=W(a^0+\theta(z)$ which lead exactly to solution of
$M_n$.

In this way we come to situation  when question of integrability is
connected with functional equation of special form. In this connection
we would like
to notice that  functional equations arises before in the theory of
integrable system for finding general solution in implicit form \cite
{FL}.$(3-4)$
What class of solutions of $M_n$ it is possible to find by this method
we don't know. What connection have all this with group theoretical
approach is unknown for us
at the present moment also. In general the very interesting problem
may be formulated as the follows it is necessary to enumerate all
functions  $W$ with choice
of which equation $M_n$ has integrable solution. Part of this
solutions in partial cases we have presented in this paper in explicit
form.
What domain of mathematic responsible for this is the most interesting
question for further investigation.

\end{document}